# Bicomp: A Bilayer Scalable Nakamoto Consensus Protocol

Zhenzhen Jiao[1,2], Rui Tian[1,3], Dezhong Shang[1], Hui Ding[1]

[1] Chaincomp Technologies Co., Ltd., Shenzhen 518000, China
[2] Institute of Computing Technology, Chinese Academy of Sciences, Beijing 100190, China
[3] Beijing Engineering Research Center for IoT Software and Systems, Beijing University of Technology, Beijing 100024, China

## ABSTRACT

Blockchain has received great attention in recent years and motivated innovations in different scenarios. However, many vital issues which affect its performance are still open. For example, it is widely convinced that high level of security and scalability and full decentralization are still impossible to achieve simultaneously. In this paper, we propose Bicomp, a bilayer scalable Nakamoto consensus protocol, which is an approach based on high security and pure decentralized Nakamoto consensus, and with a significant improvement on scalability. In Bicomp, two kinds of blocks are generated, i.e., microblocks for concurrent transaction packaging in network, and macroblocks for leadership competition and chain formation. A leader is elected at beginning of each round by using a macroblock header from proof-of-work. An elected leader then receives and packages multiple microblocks mined by different nodes into one macroblock during its tenure, which results in a bilayer block structure. Such design limits a leader's power and encourages as many nodes as possible to participate in the process of packaging transactions, which promotes the sharding nature of the system. Furthermore, several mechanisms are carefully designed to reduce transaction overlapping and further limit a leader's power, among which a novel transaction diversity based metric is proposed as the second level criteria besides the longest-chain-first principle on selecting a legitimate chain when fork happens. Security issues and potential attacks to Bicomp are extensively discussed and experiments for evaluation are performed. From the experimental results based on 50 nodes all over the world, Bicomp achieves significant improvement on scalability than that of Bitcoin and Ethereum, while the security and decentralization merits are still preserved.

## 1 Introduction

Blockchain has received great attention from both industry and academia along with the boom of cryptocurrencies in recent years. It first appears as the foundation of the famous cryptocurrency Bitcoin [1] and now has evolved to a computing platform that enables decentralized, persistent, anonymous, and auditable transaction recording at Internet scale. Although blockchain has already motivated many innovations [2-13], it still faces several obstacles which hinder its widely deployment. A famous problem is the so-called *Impossible Trinity*, which means a blockchain system cannot simultaneously achieve the following three characteristics: high scalability, high security, and full decentralization.

Specifically, in a blockchain platform, transactions are recorded by an ordered list of blocks (i.e., a blockchain). Each node in the system competes for the right of generating new blocks and the associated rewards. The process of competing and achieving a network-wide agreement among all participants called a consensus process. The consensus protocol of the first blockchain system, i.e., Bitcoin, is often mentioned as Nakamoto consensus. In Nakamoto consensus, each node is required to solve a computationally difficult problem for obtaining the right to generate a block, which is called Proof-of-Work (PoW). PoW has good anti-attack performance in open network environments and it's the only one algorithm whose security has been verified for almost ten years in the real world with millions of real users. However, its associated resource consumption is huge and its scalability is poor.

Recent emerged algorithms often aim at addressing the scalability issue by sacrificing some degree of decentralization. In this category, a famous protocol is Proof-of-Stake (PoS). The main idea behind PoS is simple: the probability to create a block and obtain the associated reward is proportional to a node's owned stake in the system. PoS is based on an assumption that users hold more stakes would be less likely to compromise the system. PoS can achieve high scalability, which attracts many systems. For example, Ethereum [14] claims to switch from PoW to PoS in the near future. However, PoS may apparently lead to centralization to a small number of users with more stakes. Similarly, Delegated Proof-of-Stake (DPoS), which is a variation of PoS and utilized by Enterprise Operation System (EOS) [15], also sacrifices the fairness for scalability and thus leads to power centralization to several users, e.g., 21 super nodes in EOS.

In this paper, we seek for achieving an equilibrium among the *Impossible Trinity* towards another direction. Our direction is to preserve PoW and its well-proven decentralization and security. On this basis, we seek methods to improve its scalability. For this purpose, let us first review the two key parameters that influence

The work in this article was supported partially by the NSF of China under Grant Nos. 61501125 and 61502018.

the scalability of Nakamoto consensus, i.e., block size and block interval. Increasing block size or reducing block interval can improve scalability, but may lead to other performance issues. For example, bigger blocks need longer time to propagate in network. Accordingly, if the block interval is not enough for a new generated block to be known by most of the network nodes, forks may happen frequently and thus lead to system instability. As a result, Bitcoin generates one 1MB block every 10 minutes and can process 7 Transactions Per Second (TPS) on average.

There already exist much efforts that aim at conquering PoW's low scalability. In [16], a novel PoW based protocol named Bitcoin-NG was proposed. In Bitcoin-NG, the process of block generation is decoupled into two phases: leader election and transaction serialization. In each epoch, a leader is elected via PoW and can continue to append transactions to the blockchain during its epoch, until a new leader is elected. Bitcoin-NG achieves throughput improvement significantly. However, a leader in Bitcoin-NG has full power to decide which transaction to be packed and thus is capable to hurt the balance of system. Furthermore, since only transactions received by the leader can be packaged, this may result in extra transmission latency or omission of remote transactions and thus affects the throughput. ByzCoin in [17] improves Bitcoin-NG by introducing a group of leaders, the so-called consensus committee, rather than one single leader. In ByzCoin, when a node solves the hash puzzle of PoW, it becomes a member of the current leader group and has the power to vote for microblocks by using a modified Practical Byzantine Fault Tolerance (PBFT) [18] algorithm. However, in ByzCoin, a new microblock must refer the previous one, which lost the possibility of concurrent packaging. Furthermore, the generation of microblocks in both Bitcoin-NG and ByzCoin only requires signature of miners, which reduce the cost of fraud. In [19], the authors combines PoW with Direct Acyclic Graph (DAG) to achieve concurrent transaction packaging thus improves the throughput. However, similar as other DAG based protocols, how to address the large latency for transaction confirmation and how to determine an effective incentive strategy are still open issues.

In this paper, we propose Bicomp, a bilayer scalable Nakamoto consensus based protocol. Bicomp addresses the scalability issues by introducing a two-phase PoW based block generation. The first phase is leader-election phase, where node performs PoW based computation to compete for leadership. Furthermore, in contrast to other work, Bicomp limits a leader's power and avoids the associated negative impact by further leveraging a PoW based microblock generation process. In this phase, a node other than a leader will package transactions it has received into a microblock, whose generation needs some PoW computation, and then sends it to the leader. A leader has no right to pack any transactions but only allowed to pack the microblocks from other nodes to generate a macroblock, which forms a bilayer structure of blocks. A leader can only produce one such macroblock during his tenure and then broadcasts it to the network. Another advantage for using microblock is its sharding nature. That is, a microblock miner can package the transactions around him, reducing the large-scale broadcast of transactions in the network. When network size is large, such manner actually partitions network nodes according to their locations and networking status. Moreover, in Bicomp, since a macroblock header with tiny size will always be broadcast first in the network to announce the beginning of a new round for block generation and attracts other nodes to follow instead of keeping mining, chain fork is relieved even when size of the final macroblock increases.

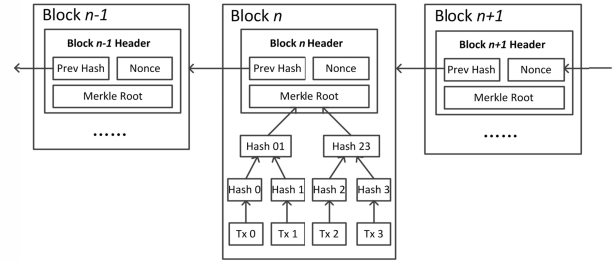

Fig. 1. Illustration to design of Bitcoin.

The following sections are organized as follows. In Section 2, we will review the existing Nakamoto consensus based protocols. In Section 3, we will provide the detailed design of Bicomp. In Section 4, security analysis is performed. In Section 5, we evaluate Bicomp by experiments. Section 6 concludes this paper.

## 2  Nakamoto Consensus Protocol and Variations

In this section, we first introduce the Nakamoto consensus protocol proposed in [1]. Furthermore, we review existing Nakamoto consensus based protocols. It should be noted that in this paper we will concentrate our attentions on PoW based protocols and its variations. For details about the other kinds of consensus protocols such as PoS please refer to [20].

**Nakamoto consensus [1]:** Nakamoto consensus is the consensus protocol of Bitcoin. In this protocol, transactions are packaged into an ordered list of blocks, with the form of Merkle tree [21]. See Fig. 1 as an example. A block is valid if its calculated cryptographic hash has $n$ leading zero bits, where $n$ is the difficulty parameter and is adjusted periodically so that new blocks have a ten-minute generation interval on average. A new generated block that has been validated will be broadcasted to the rest of the nodes. A node will append the received new block at the local stored blockchain if it verifies the correctness of the block. The protocol's security is guaranteed by the fact that the majority of the hash power in the whole system will aim at extending the legitimate chain faster than any corrupt minority that might try to rewrite history or double spend currency. However, this guarantee is only probabilistic, which leads to fundamental problems, i.e., forks and the risks of double-spending attacks and etc.

Specifically, fork may happen when multiple distinct blocks with the same parent were generated before the network reached a consensus, which is a common phenomenon when block propagation latency exists. It will take a long period that all well-behaved nodes finally agree to the longest branch chain. As a result, all transactions appearing only in the replaced block(s) become invalid and have to be re-submitted. It is apparent that the more branches appear, the more computational power will be wasted and the longer time for confirming a transaction will be consumed. Therefore, the block size in Bitcoin has to be limited to 1 MB, since a larger block size may increase the propagation delay and thus the probability of fork. Similarly, block interval is set to about ten minutes for preventing more blocks to be generated before the previous one has been propagated throughout the network. Such limitations in turn results in an upper bound on the number of TPS, i.e., an average of 7 TPS.

**Bitcoin-NG [16]**: Bitcoin-NG is the first work that divides the block generation process into two separate phases: 1) leader election, and 2) transaction packaging. To achieve this, Bitcoin-NG proposes two different block types: Keyblocks are generated through mining with PoW and are used to securely elect leaders, at a moderate frequency, such as every 10 minutes as in Bitcoin. Microblocks contain transactions, which are generated by elected leader without PoW computation. Microblocks can be produced continuously between the mining of two keyblocks, which increases the throughput. Bitcoin-NG's decoupling of keyblocks from microblocks is an important idea that motivates much research work in this area. However, a leader in Bitcoin-NG has enough power to hurt the system, i.e., a leader in his epoch can intentionally forge or rewrite history and invalidate transactions without any cost, which make Bitcoin-NG to be vulnerable to selfish or malicious manners. Moreover, since all transactions must be received by the leader, this may result in extra latency or omission of receiving remote transactions at the leader and thus affects the throughput.

**ByzCoin [17]:** ByzCoin is a novel Byzantine consensus protocol that achieves Byzantine consensus while preserving Bitcoin's open membership by dynamically forming hash power-proportionate consensus committee. Such a committee replaces the single leader in Bitcoin-NG, which is the key improvement of ByzCoin for maintaining the security of the system. In ByzCoin, when a node solves the hash puzzle of PoW, it can then become a member of the current consensus committee and has the power to vote for microblocks by using a modified PBFT algorithm. ByzCoin organizes the consensus committee into a communication tree where the most recent miner (the committee leader) is at the root. The committee leader runs PBFT to get all members to agree on the next block and it replaces PBFT's $O(n^2)$ MAC-authenticated all-to-all communication with a primitive called scalable collective signing (CoSi) [22] that reduces messaging complexity to $O(n)$. However, in ByzCoin, a malicious committee leader can potentially reject transactions by not proposing them or exclude nodes from the consensus process, which will hurt the safety and fairness. Furthermore, since a new microblock in ByzCoin must cite the previous one, its throughput is largely decreased due to the loss of parallelism.

**Conflux [19]:** The key novelty of Conflux is it allows multiple participants to contribute to the Conflux blockchain concurrently while still keeping safety of the chain. In Conflux, when a node generates a new block by PoW, it identifies a predecessor as the parent block for the new block and creates a parent edge between these two blocks. To incorporate contributions from concurrent blocks, the node also identifies all other blocks that have no incoming edge and creates reference edges from the new block to those blocks. Such reference edges represent that those blocks are generated before the new block. As a result, the edges between blocks form a direct acyclic graph rather than a chain. To determine the order of the concurrently generated blocks, Conflux establishes a pivot chain that starts from the genesis block and contains only parent edges. Further, all blocks in the direct acyclic graph are divided into epochs using the pivot chain. However, Conflux is built on an assumption that transactions in concurrent blocks will never conflict with each other and therefore cannot be used in open network environments with potential adversaries. Furthermore, similar to other DAG based systems such as IOTA [23], the incentive strategy of Conflux is not designed. Due to the openness and arbitrariness of the generation of blocks in DAG based protocols, how to achieve safe and fair incentive is still an open issue in this area.

## 3 Bicomp: A Bilayer Scalable Nakamoto Consensus based Protocol

In this section, we propose Bicomp, a bilayer scalable Nakamoto consensus protocol. We first introduce the system model. Next, we provide the design of Bicomp.

### 3.1 System Model

Bicomp is designed for open and untrustworthy networks wherein nodes may delay, drop, re-order or duplicate messages arbitrarily. A subset of system participants may be Byzantine nodes and controlled by a malicious attacker. The other nodes are honest and abide by the predefined protocol. In this paper, we assume that the total hash power of all malicious nodes is less than 1/4 of the system's total hash power at any time, since the PoW based systems are vulnerable to selfish mining attacks by adversary who controlled more than 1/4 hash power of the network [24]. Network nodes in the system are connected by reliable peer-to-peer connections. Each node can generate key-pairs but there is no trusted public key infrastructure. In addition, we assume that each node's computational power in the network is within a certain range and no one is far better than others.

### 3.2 Overview

In this section, we first provide an overview to Bicomp.

Bicomp divides the block generation process into two phases, i.e., leader election and transaction packaging, which is similar as Bitcoin-NG and ByzCoin. However, in Bicomp, PoW based computation are needed in both phases. To be specific, Bicomp employs two kinds of block, namely macroblock and microblock. The macroblocks form the main chain and each of them links to the previous one chronologically. Macroblock consists of a header and a body. The macroblock header is used to compete for leadership of the next round for macroblock generation. The generation of such a header needs a PoW based puzzle solving process. In contrast, microblock is responsible for packaging transactions. A macroblock may contain multiple microblocks which forms a bilayer structure of blocks. A microblock only belongs to one macroblock and no need to refer to any other microblocks. The structure of the two blocks and their relationship is demonstrated in Fig. 2 for clarity.

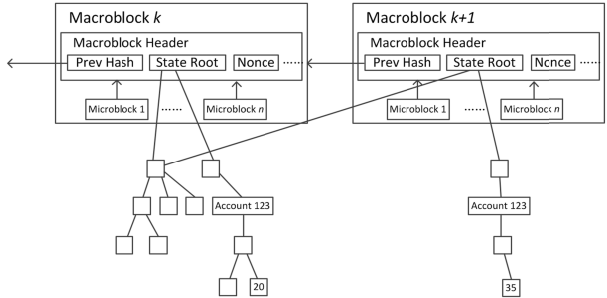

Fig. 2. Illustration to the block structure of Bicomp.

The motivations of such manner are as follows.

The first is to achieve parallelization and sharding. The propagation and transmission of a large number of transactions on a large scale is a huge burden to network. In addition, considering the existence of delays in the network, it is difficult for a single leader to collect as many transactions as possible in time before it starts to generate a new block. Those are reasons why we need parallelization and sharding. Bicomp addresses these issues by enabling transactions to be concurrently verified and packaged into microblocks by nearby nodes. As a result, transactions are packaged immediately when being generated instead of being spread through the network.

The second is to limit the power and earnings of a leader and encourage as many nodes as possible to participate into the mining process. In Bicomp, microblock miner selects transactions to package and obtains a part of the transaction fees as its reward. In contrast, macroblock miner can only package the microblocks into the macroblock and obtain the block generation reward from the system and part of transaction fees. In Bicomp, the difficulty and computing paradigm of PoW for microblock is set to be friendly to ordinary computing devices for encouraging as many nodes as possible to participate in the mining process and thus increases the diversity of microblocks. Here, the diversity of microblocks is estimated by the diversity of transactions packed in microblocks.

Bicomp in general works with the following steps:
1) Nodes perform PoW computation to compete for leadership to take charge of the next round of macroblock generation;
2) After a leader is determined, other nodes can participate in the transaction packaging process by generating microblocks and disseminating them to the leader;
3) Leader validates the received microblocks and packages them as a macroblock at the end of its tenure and broadcast it to the network;
4) Nodes who received the new macroblock will verify and append it to the local blockchain and start to compete for leadership of the next round.

The above main process is straightforward but its design and implementation is non-trivial and requires careful consideration. For example, microblocks from different nodes may contain overlapping transactions or even conflicting ones, how to deal with and also avoid such situation? How to reduce forks when block size in Bicomp becomes larger than Bitcoin's? In next subsections, we will introduce Bicomp in detail and explain how we address the issues.

### 3.3 Leadership Election via Macroblock Header

Each node in the system can compete for the leadership of a round for generating new macroblock. The node should first solve a PoW based puzzle to generate a macroblock header. Once a valid macroblock header is generated and broadcasted, it switches its role to leader and waits for microblocks from other nodes. In Bicomp, rounds for generating macroblocks are adjacent but not overlapped on the timeline, and each round is distinguished by different leaders.

Due to packet propagation latency, after a leader is already elected and within a certain zone, nodes in this zone will still receive different macroblock headers from other nodes that compete for leadership in this round. At that time, the receiving node will check the new received header and compare it with the currently held one. If the two headers refer to the same previous blocks, then the node will still use the one received before. If the two headers refer to different previous blocks, the node will know that the chain forks. A PoW based consensus algorithm often forks due to its probabilistic consistency nature [17]. Currently, fork is solved in Bitcoin and Ethereum according to the longest or heaviest principles, respectively. That is, in Bitcoin, a node always follows the longest branch chain. In Ethereum, since the existence of uncle block, a node may choose the heaviest chain to follow, where a branch chain with more blocks and uncle blocks is defined to be heavier. The main logic behind their manners is the same, i.e., the chain with higher computational effect will be reserved. In this work, we still follow such idea but with a slight modification since there exists two kinds of PoW based computational work in Bicomp, whose computational difficulties are different. To avoid the selfish mining attack that a malicious leader increases the weight of its selfish mined chain by adding multiple microblocks, whose mining difficulty is often less than macroblock, we define a mechanism to choose one from branches as follows. A chain with more macroblocks will be chosen when they have different numbers of macroblocks; otherwise, the chain whose embedded microblocks contained more non-overlapping valid transactions will be chosen. The benefits for such manner are manifolds. First, the effect of macroblock mining is taken into account. Second, selfish mining on microblock is forbidden. Third, it encourages the leader to choose more non-overlapping transactions when it has choices.

Except for solving a fork when it appears, Bicomp also suppresses the occurrence of forks by the existence of macroblock header and its spread at the beginning of each round. The reason is, the size of macroblock header is only 200 bytes in Bicomp. Consequently, less time will be consumed for spreading it throughout the network, comparing with the 1 MB block of Bitcoin and 2KB block of Ethereum. When a node receives a header, it will stop competing for leadership and wait for the incoming macroblock, by which the potential forks are suppressed.

Another instant a node may encounter a fork is when the node mines a macroblock header. In Bicomp, a node must refer to the last macroblock's hash value before it starts PoW based header mining. However, once a node receives a different chain from the one it maintains, even if it has successfully generated a macroblock header and becomes a leader, it will still give up its leadership and switches to the newly received chain when the latter one should be legitimate, in order to maintain the consistency of the blockchain.

Besides, there exists a key step before a node starts to mine a macroblock header. That is, a node should calculate, verify, and settle the transactions contained in the macroblock it wants to refer and also the associated fees and rewards. This is a key step which is helpful to guarantee the system security and avoid potential fraud of a leader. We will introduce this part and explain in detail in section 3.6.

### 3.4 Packaging Transactions in Microblocks

Microblock mining is the process that node packages transactions into microblocks it generates. Multiple nodes can mine microblock concurrently in Bicomp and most of their efforts will be reserved, which is very different from traditional PoW based mining where only one block will be reserved at last.

In Bicomp, after a leader is elected via the mining process of macroblock header, it will broadcast its header to others. Other nodes that received the header can start to mine a microblock if they verify the credibility of the information contained in the header. In general, the verification process includes:
1) Check the validity of information in header, such as height, timestamp, and especially the correctness and identity of the hash value of the referred previous macroblock;
2) Check the correctness of the settlement results of fees done by the header, by reviewing and recalculating the related transactions based on locally stored blocks and information.

A node will accept the leadership of another node only when it has validated the correctness and legality of the macroblock header sent by that node. When a node accepted a leader, it then starts the microblock mining process and becomes a microblock miner.

A microblock miner selects transactions it has received and packages them as a Merkle tree [21]. In the Merkle tree, the leaf node stores the transaction content while the intermediate node stores the hash value of the content. The hash value at the tree root is inserted into the header of the microblock.

Different microblock miners may select the same transactions to package into their microblocks and thus lead to overlapping. To encourage microblock miners to select transactions with low possibility to be overlapped with the ones packaged by other miners, incentives are design that when a transaction is overlapped with its prior ones, it will be considered non-existent and its miners cannot get any reward. Therefore, each microblock miner is motivated to select transactions that avoid confliction with others. For example, a miner can leverage machine learning or game theory based localized methods, to determine a selected transaction set.

Furthermore, a networking based mechanism is leveraged by Bicomp to help reduce such overlapping. That is, in the underlying P2P network, a transaction is only relayed within limited hops. Such manner takes advantage of the sharding nature of network and reduces the possibility that the same transaction to be received by many different microblock miners.

PoW based computation is the last step to generate a microblock. The microblock miner continuously changes the nonce until the calculated hash value meets the difficulty requirement. Information including the chosen transactions are involved in calculation, which aims at increasing the fraud cost of a malicious miner. When a verifiable microblock is produced, the miner will then broadcast it in the network. Next, a miner can start to mine another microblock until it receives the macroblock generated by the leader of this round.

### 3.5 Macroblock Generation

Macroblock is generated by the elected leader at its tenure and will also ends his tenure when finished.

After a leader mines a macroblock header successfully as introduced in section 3.3, it waits and receives newly generated microblocks from other nodes. After verifying the microblocks, leader puts them into its caching pool.

Each leader's tenure lasts for $T$ minutes, which starts at the moment when that leader sends out its mined macroblock header, and ends at the time when it picks microblocks from the caching pool and packages them together to generate a macroblock. The leader signs the generated macroblock and broadcasts it to the network. In Bicomp, a leader always waits a fixed amount of time $T$ before it starts to generate a macroblock, thus different numbers of microblocks may be received at different rounds. In general, two cases will exist as follows.

The first case is, when time $T$ elapses, the microblocks received by the leader do not exceed a predefined capacity $C$. In this case, all these microblocks are adopted and encapsulated into the new macroblock, unless there is no transaction in a microblock that is not included in previous ones.

The second case is a leader receives more microblocks than it can accommodate, and it will pick some to package. For a non-malicious node, it will always select the microblocks whose transactions inside has lower transaction overlapping number, so that not only the node obtains higher rewards by itself but the microblock achieves a higher probability for being chosen as part of the legitimate chain when chain forks, as we discussed above.

An exception will happen when the leader is a malicious node. Since no other node can know exactly which and how many microblocks the leader has received, so that the leader has the ability to arbitrarily choose microblocks and thus the transactions therein. In such situation, our mechanism discourages node to do so because it cannot achieve extra benefit except undermining the system. However, such manner cannot be completely prohibited. Fortunately, due to the mechanisms such as election and the new defined metric in Bicomp for selecting branch chains, a malicious node cannot always achieve its purpose but has high possibility for paying high cost for its behavior.

Another important problem is how to avoid the double-spending risk by the leader. Think about the scenario that a leader generates two macroblocks. It can disseminate the first one with a prepared transaction *TX* in the network. After a while, when others believe the macroblock and transaction *TX* have been confirmed, the leader then replaces the macroblock with the second one without *TX*, and thus performs a double-spending attack. We will discuss this issue in 4.2.

### 3.6 Incentives and Confirmation

Incentives in Bicomp aim at not only motivating nodes to join the block mining process, but also encouraging miners, for both macroblock and microblock, to select non-overlapping transactions. Before explaining this, we shall first introduce how to decide a transaction's validity when it conflicts or overlaps with another one that located in another microblock but also packaged in the same macroblock.

Recall that in Bicomp, the microblocks contained in a macroblock may have overlapping transactions due to concurrent transaction packaged by different nodes in a pure open network environment. Therefore, in some cases, a macroblock may have stored the same transaction several times, or even conflicting transactions. If the overlapped or conflicted transactions are allowed to be executed at the same time, a double-spending attack might occur, so we must avoid such situation. Therefore, in Bicomp, the validity of a transaction in a macroblock depends on its order, i.e., when a prior transaction is confirmed, a later one will be considered as invalid if it conflicts or overlaps with the prior one. However, the above dependency on placing order may enlarge a leader's power and further makes the chain vulnerable to Sybil attack [26]. For example, in Sybil attack, a leader can determine the orders of the transactions by choosing and sorting microblocks from its companions according its own benefit. This issue also exists in all existing work that uses leader to perform transaction packaging. However, in Bicomp, the existence of the microblocks helps reducing such risks: First, a leader can only choose microblocks but not select transactions directly as done in other work. This introduces non-negligible computational burden to malicious nodes since microblocks need to be mined using PoW. Second, transaction diversity plays a vital role in several aspects in Bicomp, which also largely reduces the risks from Sybil attack since a selfish-organized mining pool tends to have lower transaction diversity than purely open network environment.

Next, we discuss how to avoid the rate of transaction overlapping by using incentives in Bicomp. As we introduced above, several mechanisms are designed to relieve overlapping, such as limiting transaction dissemination range in network, and using the transaction diversity as a metric to estimate the forking chain's weight. Beside these mechanisms, the incentives mechanism is another effective and straightforward tool for achieving the goal.

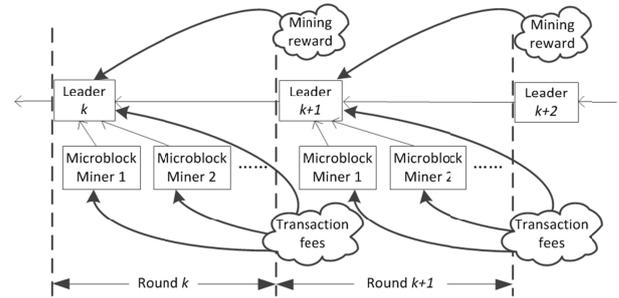

Fig. 3. Illustration to the incentives design in Bicomp.

In Bicomp, incentives are performed based on token/cryptocurrency, which come from two sources, i.e., block mining reward and transaction fee. The block mining reward is only paid to motivate nodes to compete for leaders and mine for macroblocks. Transaction fee is the gas paid by each transaction to the system. In our mechanism, a leader can earn mining reward of macroblock. Besides, it can also obtain a percentage of transaction fees from the confirmed valid transactions packaged inside its microblocks. In contrast, microblock miner only earns transaction fees from the packaged transactions that are confirmed to be valid. As introduced in the last paragraph, a transaction will be confirmed as valid only when there exists non-overlapping and conflict transactions prior to itself. Thus, the most effective strategy to maximize a miner's income is selecting the transactions that have low possibility to collide with others.

The aforementioned incomes of both kinds of miners occurred in a macroblock will only be settled before the generation of the next macroblock header. As illustrated in Fig. 3. This is a mechanism designed in Bicomp to limit a leader's power and possibility to attack the system. Specifically, when a node wants to generate a macroblock header for electing as a leader, it will firstly verify and settle all the transactions and rewards occurred in the macroblock it wants to refer to by refreshing the state tree of accounts. Next, it inserts the root of the state tree into the prepared header and starts a PoW computation. Due to such design, any mistake or malicious behavior of the prior leader will cause no one referring to his macroblock. As a result, the malicious leader will obtain no actual incomes. Furthermore, the microblock miners who want to follow a leader will also verify the correctness of the leader's settlement for previous macroblock, to avoid malicious behaviors of the leader new elected.

## 4 Security Analysis

In this section, we provide a security analysis to Bicomp. We will discuss typical issues including selfish mining, double-spending risk, fork, and single-point failure as follows.

## 4.1 Selfish mining

In a selfish-mining attack, attackers keep discovered blocks private and intentionally fork the chain. In Bicomp, selfish miners might either detain discovered macroblock headers or keep the constructed macroblocks private. While in Bicomp different macroblock containing the most non-overlapped transactions will be chosen as the legitimate one, selfish miners may have lower chance to develop a private branch contains more non-overlapped transactions than the public branch since keeping macroblock header private prohibits other nodes to mine microblocks on it. As a result, detaining a macroblock header will not be enough to perform a selfish mining attack.

Another choice is to retain the constructed macroblocks, which is similar as Bitcoin. Bicomp demands the leader to send out macroblocks at the end of its tenure, which implies that the selfish miner cannot always retain a macroblock, because this may cause other nodes to discard its leadership. Instead, in Bicomp, an attacker can choose to construct a macroblock in advance and keep it private until its tenure ends, so that it can mine the next macroblock header with a longer time than other nodes. This behavior is consistent with the analysis in literature [24], so the conclusion therein also stands here. That is, adversary must control more than 1/4 hash power of the network to perform such attack. However, we should note that additional obstacles are set up by Bicomp, i.e., the above behavior will likely in turn make the hidden macroblock contain less non-overlapped transactions than normal public ones. This implies extra cost and risk to an attacker since if there exist multiple macroblocks in the network. In that case, the selfish mined ones are likely to be discarded by honest nodes.

## 4.2 Double-Spending

A double-spending attacker may first publish a transaction $TX_1$. When he receives a service or a commodity from the seller, he publishes an alternative transaction $TX_2$. In Bitcoin, blocks are infrequent and miners collect transactions until they form a block. Until that time, transaction $TX_1$ may be replaced by another transaction $TX_2$ without any cost. Therefore, according to an empirical conclusion, when a transaction is packaged into a block and follows by 6 new blocks later on a chain, that transaction can be considered to be confirmed. Furthermore, publication of conflicting transactions with different destinations is prohibited by standard Bitcoin software, which also warns the user of conflicting transactions propagating in the network [27]. In Bicomp, as introduced in section 3.5, each macroblock that contained transactions is referred to by the following macroblock on the blockchain. Therefore, to perform a double-spending attack, a node must have enough hash power to realize a selfish mining attack as follow.

A selfish attacker constructs two different macroblocks using the same header privately: *M1* and *M2*. *M1* contains less non-overlapped transactions but has a poison transaction $TX_P$. *M2* contains more non-overlapped transactions but without $TX_P$. The attacker first reveals *M1* to public and privately mine on *M2*. In this situation, even if honest miners construct a new macroblock on *M1*, say *M1'*, the private attacker can still use *M2* and the selfish mined *M2'* to replace both *M1* and *M1'* on the legitimate chain. Since *M2* contains more non-overlapped transactions than *M1*, this may increase the possibility of success with such replacement. Such behavior can result in double-spending attacks. However, we should note that *M2* is also a selfish mined block which may have low diversity compared to the public-mined one, i.e., *M1'*. Furthermore, to realize such attack, at least 1/4 hash power is required even in Bitcoin as introduced above. Due to the existence of multiple defensive mechanisms in Bicomp as introduced, it is intuitive that more hash power is required to perform such attack successfully in Bicomp. We leave the theoretical proof as a future work. Besides, a delay confirmation for transactions as done in the Bitcoin is also useful to avoid such attack in Bicomp.

## 4.3 Fork

Fork in Bitcoin or other PoW based protocols often happens due to the existence of packet propagation delay in network. That is, multiple blocks are generated after the same prefix since the miners fail to receive a previous generated block before they repeat the mining work. To solve this, Bitcoin limits the block size to 1 MB.

In contrast, Bicomp decreases the chance of such fork since a macroblock header will always come first before the generation of other kind of blocks. Therefore, microblock miners will always choose a header to follow before it starts to mine and wait for the incoming macroblocks. Recall that a header's size is only 200 bytes, so that it can be spread throughout the network very quickly and thus largely decrease the possibility of fork.

However, in a large network, Bicomp may still experience fork when multiple macroblock headers are generated after the same prefix of a macroblock due to the network delay. We claim that such fork is not purely harmful to our system. That is, as we have discussed, a branch with more computational power and high transaction diversity will be selected as the legitimate one, the existence of fork has the potential to curb the effects of Sybil attacks and selfish mining.

## 4.4 Single-Point Failure at Leader

In Bicomp, there exists only one leader who is allowed for mining a macroblock at one round. As a result, mechanisms should be designed to avoid single-point failure at leader and resist to Distributed Denial of Service (DDoS) attack. In our design, each node who has accepted a macroblock header from a leader, will wait until it receives a valid macroblock or the end of that leader's tenure, whenever comes first. If a node never receives a valid macroblock from the current leader before its tenure ends, it will initiate the leader-election process based on the latest macroblock restored locally. Here, those headers, which were generated at the same time with the out-of-date one, can be reused and re-broadcasted immediately to the network. In terms of judging a leader's expiration locally, a node *b* should determine a waiting time *E* as follows:

$$E = T(a) - [T_b(a) - T_{ts}(a)] + \delta, \quad (1)$$

where $T(a)$ is the tenure length of the leader $a$, which is calculated according the latest macroblock header's mining difficulty and can be obtained by nodes locally. $T_b(a)$ is the time node $b$ received the header from $a$. $T_{ts}(a)$ is the timestamp when the header is generated at $a$ and is carried by the header. $\delta$ is an estimated time based on the network size, which is used to measure the transmission time of the final generated macroblock from the leader $a$ to node $b$. In order to prevent a header from using a fake timestamp to gain extra time for itself, such as using an earlier one than real, the local node should judge its authenticity and legality based on local knowledge. These knowledge may include the generation time of previous macro- and microblocks, the expected time for computing a header based on the current calculation difficulty, timestamps from neighbors, and timestamp server similar as used in [1].

## 5 Evaluation

In this section, we investigate the performance of Bicomp on a test network constituted by 50 nodes, whose geographical distribution is shown in Fig. 4. These nodes are cloud servers belonging to different cloud service providers, such as Alibaba Cloud and Vultr. Each node has 2 cores (2.6 GHz frequency each), 8 G byte memory, and 1Gbps network bandwidth. By default, in our experiment, transactions are generated randomly in the network at the rate of 1000 per second. Each microblock can contain up to 1500 transactions. The transmission range of each transaction is limited to two hops. With the computing power of current CPUs, the defined difficulties allow one macroblock header or microblock generated per minute approximately.

### 5.1 Throughput

We first evaluate the throughput in terms of TPS that Bicomp can achieve. Furthermore, we evaluate the number of non-overlapped transactions packaged in a macroblock, by enabling each microblock miner to select transactions randomly when packaging its microblock. We vary macroblock capacity, which is the upper bound of number of microblocks in a macroblock, to reveal its impact on throughput under different macroblock intervals, i.e., 60, 120, and 180 seconds, respectively. Note that the macroblock interval also is the tenure length of a leader.

In Fig. 5, we can find that Bicomp always achieves high throughput in different macroblock intervals, which varies within [160, 507] when counting all transactions and [113, 371] when only non-overlapped transactions are recorded. Also, shorter macroblock interval leads to higher throughput at the same

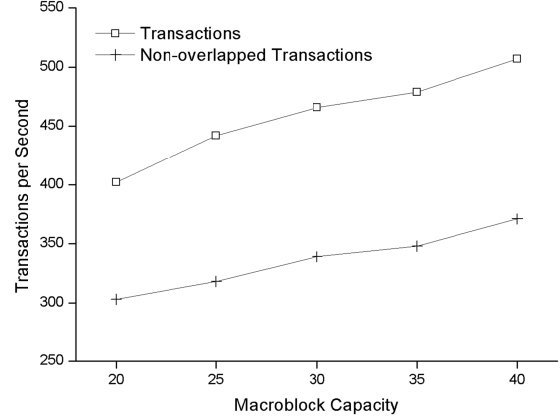

(a) Macroblock interval of 60s

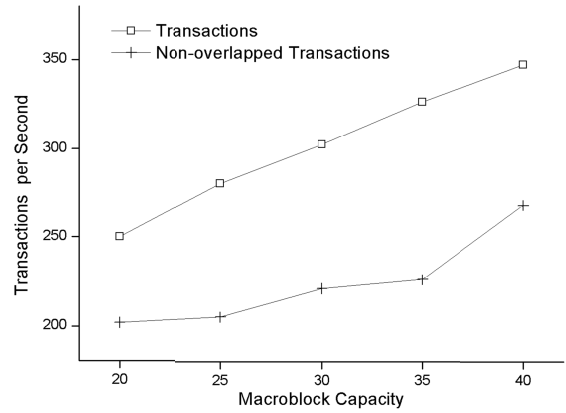

(b) Macroblock interval of 120s

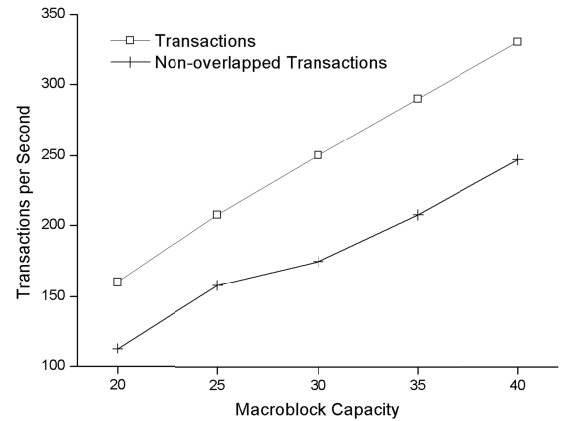

(c) Macroblock interval of 180s

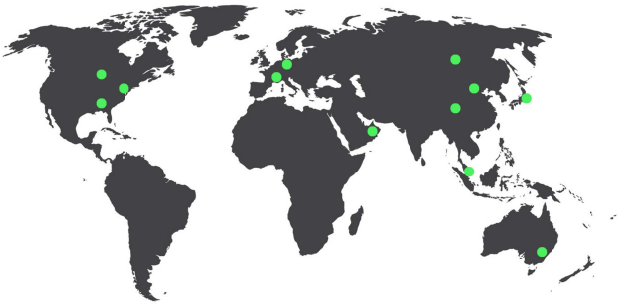

Fig. 4. Distribution of test nodes.

Fig. 5. Influence of the macroblock capacity on transaction throughput.

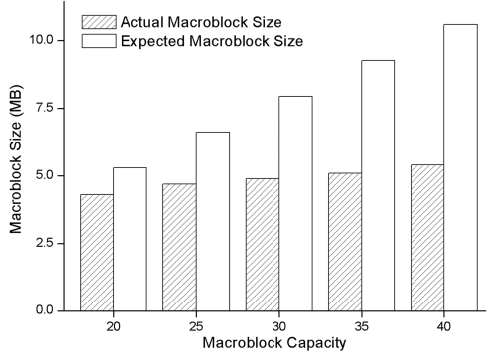

(a) Macroblock interval of 60s

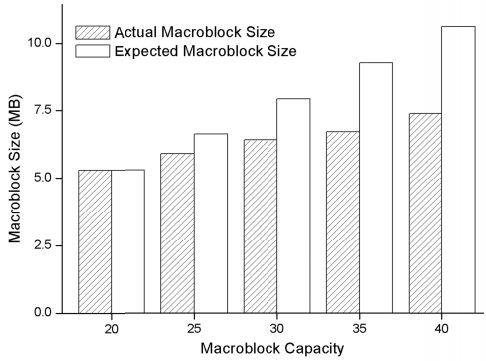

(b) Macroblock interval of 120s

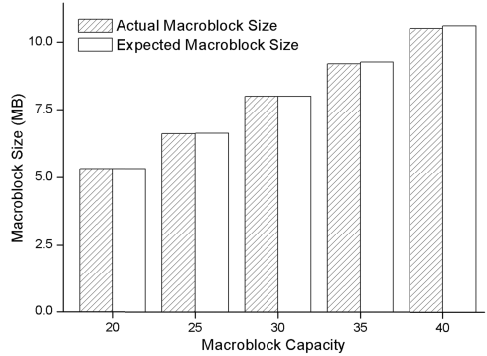

(c) Macroblock interval of 180s

Fig. 6. Influence of the Macroblock Capacity on Macroblock Size.

macroblock capacity, which is consistent with the expectations. However, in Bicomp, a short interval does not suitable for most situations and performance requirements. We can further find in Fig. 7 that a shorter interval may lead to a slower transaction confirmation, which is conflicted to common sense. We will explain this in the following paragraph. In Fig.5, when macroblock interval is fixed, the throughput improves by increasing the macroblock capacity. However, when comparing data across Fig. 5(a), 5(b) and 5(c), we can also find that the throughput gain brought by increasing macroblock capacity when macroblock interval is short, is less than that when macroblock interval is long. This is because, in the case with shorter interval, a microblock miner cannot receive enough non-overlapping transactions to fill diverse microblocks, and thus lead to high overlapping rate.

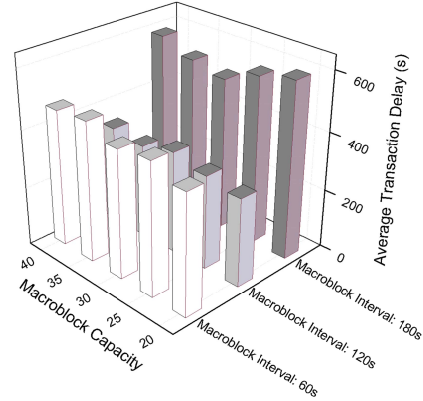

Fig. 7. Average delay for packaging transactions.

### 5.2 Block Size

In Fig. 6, we compare the average macroblock size (in byte), i.e., actual one with expected one, versus different numbers of microblocks in a macroblock. Here, the actual average macroblock size is the average size of actual generated macroblock in each round. The expected size of a macroblock is the pre-defined upper bound of the number of microblocks in a macroblock. In Fig. 6, shaded columns illustrate the average actual size of macroblock and the white ones illustrate the expected macroblock sizes.

From the figures, we can see that prolonging the macroblock interval can increase the number of microblocks packaged in each macroblock. However, as observed in Figs. 5 and 6, shorter interval, such as 60 seconds, can result in higher TPS and smaller macroblock size and thus a lower propagation burden to network, which would be a better option than the longer ones.

### 5.3 Average Latency for Packaging Transactions

We evaluated the average latency for transaction packaging under different settings as introduced above. In Fig. 7, we can find that the lowest average latency for packaging transactions is obtained when macroblock interval equals to 120 seconds. The reason is the 60-second-interval is not long enough for a leader to receive sufficient microblocks, which is also indicated by Fig. 6 (a). Furthermore, considering that 1000 transactions per second being continuously injected into the network, a 180-second interval may result in 1.8E+5 new transactions added in one interval, which may cause most transactions being backlogged, and thus induce longer latency than the case with shorter intervals.

### 6　Conclusion

In this paper, we propose Bicomp, a bilayer scalable Nakamoto consensus based protocol. Bicomp introduces a two-phase block generation process, i.e., leader election and transaction packaging. In the leader-election phase, PoW based computation is employed for generating a macroblock header to compete for leadership. In transaction packaging phase, any node except the leader can join the process and package transactions they received and generate microblocks, which compose the final macroblock. In Bicomp, a leader has no right to pack any transactions but only allowed to use microblocks from others to generate a macroblock, which forms a bilayer structure of blocks. By doing so, the leader's power is limited and other nodes are also encouraged to play significant roles in block generation process. Besides, incentives are also elaborately designed to motivate different participants. In general, Bicomp fairly decentralizes network power, utilizes the sharding nature of the network, increases the efficiency via concurrent packaging process, and reduces the long-distance transmission and flooding of numerous transactions in the network. Security analysis and evaluation for Bicomp are proposed, which shows better performance than Bitcoin and Ethereum in terms of the security and scalability.